\documentclass[fleqn,usenatbib]{mnras}
\usepackage{newtxtext,newtxmath}   % 字体设置优先加载
\usepackage[T1]{fontenc}           % 字体编码紧随其后
\usepackage{graphicx}              % 图形支持
\usepackage{amsmath}               % 数学环境
\usepackage{threeparttable}        % 表格注释
\usepackage{booktabs}              % 表格样式增强
\usepackage{float}                 % 浮动体控制
\usepackage{subcaption}            % 子图（依赖float或caption宏包）
\usepackage{url}                   % URL处理
\usepackage{hyperref}              % 超链接
\usepackage{caption}
\usepackage{comment}

\title[The energy structure function of FRBs]{The energy structure function of fast radio bursts supports a stochastic origin model}

\author[Chen, Zhang, \& Dai]{
Yi-Nan Chen$^{1,2}$,
Yong-Kun Zhang$^{3}$,
Zi-Gao Dai$^{1,2}$
\thanks{daizg@ustc.edu.cn}
\\
$^{1}$Department of Astronomy, University of Science and Technology of China, Hefei 230026, People’s Republic of China\\
$^{2}$School of Astronomy and Space Science, University of Science and Technology of China, Hefei 230026, People’s Republic of China\\
$^{3}$National Astronomical Observatory, Chinese Academy of Sciences, Beijing 100101, People’s Republic of China}

\date{Accepted XXX. Received YYY; in original form ZZZ}
\pubyear{2025}

\begin{document}
\label{firstpage}
\pagerange{\pageref{firstpage}--\pageref{lastpage}}
\maketitle

\begin{abstract}
The origin of fast radio bursts (FRBs) has remained a mystery up to now. There are two kinds of process invoking neutron stars as an origin of FRBs, namely inner-driven starquakes and outer-driven collisions with interstellar objects (ISOs). The former origin should exhibit an earthquake-like statistical behavior while the latter should show a stochastic process. In this paper, we introduce a new statistical method by making use of the energy structure function of active repeating FRBs and earthquakes. We find that the energy structure function of FRBs exhibits a very different statistical behavior compared to that of earthquakes. On small time-interval scales, the energy of an earthquake show a tendency to decay with time-interval and the energy difference of a pair of events increases with time-interval. Such a behavior is not found in FRBs, whose energy function is very similar to those of a stochastic process. Our result shows that repeating FRBs may have an origin process differing from that of earthquakes, i.e., FRBs arise from a series of unrelated events such as collisions of a neutron star with ISOs.
\end{abstract}


\section{Introduction}
Fast radio bursts (FRBs) are transient radio phenomena characterized by millisecond durations and extremely high brightness temperatures (\(T_B \sim 10^{36}\) K), whose physical origin has remained enigmatic since their discovery in 2007 \citep{Lorimer2007}. The dispersion measures (DMs) of these bursts typically far exceed contributions from the Galactic interstellar medium, indicating a cosmological origin. Based on their activity patterns, FRBs are categorized into repeating bursts and non-repeating bursts. More than 60 FRBs have been reported to repeat \citep{Spitler2016,chime2019a,chime2019b,Kumar2019,Luo2020,Niu2022}. 

It remains unclear whether repeating and apparently non-repeating FRBs share a single physical mechanism. Statistical analyses have revealed differences in observed parameters between the two classes \citep{chime2023ApJ}, which may point to distinct origins. However, analyses of activity rates suggest that both types of FRBs could arise from a common source \citep{250609138B}. A crucial observational clue comes from the association of FRB 20200428 with the Galactic magnetar SGR 1935+2154, which offers valuable insights into the possible origin of FRBs \citep{Bochenek2020, chime2020, Mereghetti2020}. Although the energy of FRB 20200428 is significantly lower than that of cosmological FRBs, this event demonstrates that magnetars are capable of producing FRB-like bursts. While the physical origin of FRBs remains uncertain, magnetars such as SGR 1935+2154 may be responsible for at least a subset of FRB events.

The generation of FRBs in a magnetar environment requires a triggering mechanism \citep{2023RvMP...95c5005Z}. The existing models include starquakes \citep{Wang2018}, interstellar object (ISO) collisions \citep{Geng2015,Dai2016}, spontaneous magnetic reconnection events \citep{Popov2010}, and vacuum gap discharge events \citep{Katz2017}. Currently, the first two models have been widely discussed. The starquake model posits that FRBs are powered by violent activities of highly magnetized neutron stars (magnetars). In this scenario, stresses accumulated in the neutron star crust—due to magnetic field evolution or spin-down—exceed a critical threshold, leading to abrupt crustal fractures (starquakes) \citep{Thompson2001, Beloborodov2007, Wadiasingh2019, Dehman2020, Yang2021}). Alfv\'{e}n waves generated by crustal oscillations propagate into the magnetosphere, triggering coherent radiation processes via magnetospheric plasma instabilities. Recent studies have shown that during stellar oscillations, fast magnetosonic waves may also be generated, which is one of the possible mechanisms for the formation of FRBs \citep{Zhang2022,Vanthieghem2025,250813419,250812567}. In addition, the magnetic pulse generated during starquakes can also produce FRBs through forced magnetic reconnection \citep{Lyubarsky2020}.

The ISO collision model attributes FRBs to interactions between neutron stars and ISOs such as asteroids. This model encompasses two mechanisms: (1) The direct impact model suggests that the gravitational potential energy of an ISO is converted into radio emission during the infall of debris into the magnetosphere \citep{Geng2015, Dai2016}. (2) The indirect interaction model proposes that Alfvén wings—structures formed by the interaction between a neutron star's wind and an ISO in a close orbit—destabilize magnetospheric currents, producing coherent radiation without requiring ISO destruction \citep{Mottez2014}. Based on the ISO collision model, the phenomenon of FRBs associated with X-ray bursts is also well explained \citep{Dai2020}.

By analyzing the observational data through various methods, some studies favor the starquake model while others support the ISO collision model. Regarding the starquake model, soft gamma-ray repeaters (SGRs) were found to be statistically associated with earthquakes \citep{Cheng1996}. \citet{Totani2023} suggested that both the waiting time distribution of FRBs and earthquakes deviate from a Poisson process, though the specific patterns of deviation differ.
The cumulative energy distribution of FRBs also exhibits behavior reminiscent of earthquakes \citep{Wu2025}, yet using the cumulative distribution for analysis may lead to the loss of certain information. 
These similarities are not sufficient to prove that FRBs do indeed originate from starquakes.

Some works prefer the ISO collision model. By using a model that links the properties of planetesimals to FRBs, \citet{Pham2024} showed that the distributions of duration and energy of FRBs are consistent with the observed size distributions of Solar System planetesimal populations. This agreement with observed Solar System populations enhances the credibility of the ISO collision model. In the bi-variate time-energy domain, the behaviors of repeating FRBs deviate significantly from those of earthquakes. FRBs exhibit a level of stochasticity far exceeding that of earthquakes. The pronounced stochasticity may arise from the combination of diverse emission mechanisms/sites \citep{Zhang2024}. 

So far, most statistical studies comparing FRBs and earthquakes have focused on simple comparisons of individual-parameter distributions, such as similarities and differences in waiting-time or energy distributions. In contrast, there have been relatively few systematic efforts that use more advanced statistical methods to model multiple parameters jointly, thereby providing a more integrated perspective on their emission mechanisms. In this paper, we introduce a new statistical moment—the energy structure function (ESF)—which quantifies differences and fluctuations in burst energy across time scales. We apply the ESF to FRBs and earthquakes to probe their commonalities and distinctions. Because the ESF requires extensive, time-resolved datasets, we draw on the high sensitivity and sustained monitoring provided by the Five-hundred-meter Aperture Spherical radio Telescope (FAST), which has enabled large FRB samples and detailed energy distributions for repeating sources, exemplified by the 1,652 bursts detected from FRB 20121102A \citep{Li2021}. Accordingly, our FRB analysis is based on FAST observations.

This paper is organized as follows. In Section \ref{sec:2}, we list the data sources used and introduce an energy structure function . In Section \ref{sec:3}, we present the results of an energy structure function. In Section \ref{sec:anl}, we analyze the underlying physical causes responsible for the observed statistical results and discuss the statistical completeness. Finally, we provide a further discussion based on relevant works in Section \ref{sec:5}.

\section{Data and Method}
\label{sec:2}

\subsection{Data}

To compare the energy structure functions of FRBs and earthquakes, we selected well-observed datasets with high completeness.

For FRBs, we chose several sources with long-term monitoring observations, including FRB 20121102A \citep{Li2021}, FRB 20201124A \citep{FRB201124}, FRB 20220912A \citep{FRB220912}, and FRB 20240114A \citep{Zhang2025arXiv}. These datasets span periods of several months or longer and were obtained with the high sensitivity of FAST, providing more complete burst sequences than those from other telescopes.

For earthquakes, we primarily used data from Southern California in 2019\footnote{\url{https://www.fdsn.org/networks/detail/CI/}}, obtained from the SCSN catalog, and included the 2011 Tohoku earthquake\footnote{\url{https://www.usgs.gov/natural-hazards/earthquake-hazards/lists-maps-and-statistics}} from the USGS dataset for comparison (related comparative figures are shown in the appendix).

On July 4, 2019, a magnitude 6.4 earthquake struck southern California in the U.S. We select this event as the main shock due to its significant magnitude relative to recent seismic activity in this region over the last decade. For the comparative analysis that follows, we need to perform some preprocessing on the original dataset. The main shock of the earthquake occurred at 117.5$^{\circ}$W and 35.7$^{\circ}$N. We select earthquake data in the range 116.5$^{\circ}$W-118.5$^{\circ}$W, 34.7$^{\circ}$N-36.7$^{\circ}$N. The time span is $6\times10^6$ s (about 70 days) after the main shock. The choice of time span does not affect our results and the robustness is proved in Section \ref{sec:sel}. The energy of an earthquake was calculated from the magnitude by the formula\,$E=10^{11.8+1.5M}$ erg, where M is the Magnitude. We filter the data using $10^{15}$ erg, $10^{16}$ erg and $10^{17}$ erg as the lowest energy thresholds, of which the corresponding magnitudes are 2.1, 2.8, and 3.4, respectively to simulate the possible absence of low-energy earthquakes, and deposited them into three datasets: data1, data2, and data3. Moreover, to specifically investigate the statistical influence of the main shock, we select a period of data spanning the same duration but shifted by $10^6$s relative to data1 to creat data4. The main shock of the Tohoku earthquake occurred at 142.6$^{\circ}$E and 38.1$^{\circ}$N. We select earthquake data in the range 141.6$^{\circ}$E-143.6$^{\circ}$E, 37.1$^{\circ}$N-39.1$^{\circ}$N and applied similar processing to these data.

To serve as a stochastic reference for comparison with FRBs and earthquakes, we generate a synthetic dataset of 10,000 events randomly distributed in the two-dimensional time–energy space. Time values are sampled from a uniform distribution over the interval $0$ to $6\times10^6$ s (the same time span as an earthquake), while energy values are independently sampled from a uniform distribution spanning $10^{15}$ to $10^{19}$ erg.

\subsection{Energy structure function}

The structure function is a statistical moment used to describe the fluctuation of physical quantities in a field. Initially, in fluid dynamics, the velocity structure function is commonly employed to characterize velocity differences across varying spatial separations within fluid flow fields \citep{Kolmogorov1941}. Its definition is the average of the $n$th power of the velocity difference between any pair of spatial points that are separated by $l$ as given by the following formula
\begin{align}
    S_n(l) = \left< (v(x+l)-v(x))^n \right>,
    \label{eq1}
\end{align}
where $\left<\cdot\right>$ is averaging over all pairs of points that satisfy the interval $l$. As the theory continues to evolve, the second-order structure function $S_2(\tau)$ has been a popular statistic in many fields for characterizing variation across time or space. In astrophysics, it was employed to study the light curves of active galactic nuclei \citep{Kozlowski2016ApJ,DeCicco2022}. In geophysics, it was used to study the characteristic time scale of auroral electrojet data \citep{Takalo1994}. It was also used in the research of rotation measure and dispersion measure of FRBs \citep{Mckinven2023,Yang2023,Shaw2024arXiv,Li2025}.

Following the form of equation \ref{eq1}, we define a new statistical moment named the energy structure function to characterize temporal correlations across different timescales in time series. We define the first-order and second-order energy structure functions as follows
\begin{align}
    {S_1(\tau) = \left< E(t+\tau)-E(t) \right>.}
\end{align}
\begin{align}
    S_2(\tau) = \left< [E(t+\tau)-E(t)]^2 \right>,
\end{align}

According to the definition, the first-order structure function characterizes the average rate of energy variation over time scale $\tau$. The sign indicates the direction (decay or increase) and the magnitude reflects the average size of this energy change. The second-order energy structure function characterizes the average energy difference between events with a time difference $\tau$. A larger value of the function implies that the difference is greater, while a smaller value of the function indicates that the energies of two events are closer to each other. 

In addition to visually characterizing the variability between events, the structure function can also be effective in describing the correlation between event signals by demonstrating the relationship linking the structure function to auto-correlation functions \citep{Press1992}. For a stationary process without trends, the second-order energy structure function can be expanded by
\begin{align}
    S_2(\tau) &= \langle [E(t+\tau) - E(t)]^2 \rangle \nonumber \\
    &= \langle \{[E(t+\tau) - \mu] - [E(t) - \mu]\}^2 \rangle \nonumber \\
    &= \langle [E(t+\tau) - \mu]^2 + [E(t) - \mu]^2 - 2[E(t+\tau) - \mu][E(t) - \mu] \rangle \nonumber \\
    &= 2\sigma^2 - 2\sigma^2 R(\tau),
    \label{eq4}
\end{align}
where $\mu$ is the mean energy of this time series, $\sigma$ is the standard deviation and the auto-correlation function $R(\tau)$ is
\begin{align}
    R(\tau) &= \frac{\langle [E(t+\tau) - \mu][E(t) - \mu]\rangle}{\sigma^2}.
\end{align}

We know that the closer the auto-correlation function is to zero, the weaker the auto-correlation of the signal is. We can derive that
\begin{align}
    R(\tau) &= 1 - \frac{S_2(\tau)}{2\sigma^2}.
\end{align}

For a non-stationary process, the energy can be decomposed into a time-varying mean component and a fluctuation term,
\begin{align}
    E(t) &= \mu(t) + \epsilon(t),
\end{align}
where $\mu(t)$ is the time-dependent mean energy, and $\epsilon(t)$ is the fluctuation term relative to the mean energy at time \(t\). The second-order energy structure function is expanded by
\begin{align}
    S_2(\tau) =& \langle [\mu(t+\tau) - \mu(t) + \epsilon(t+\tau) - \epsilon(t)]^2 \rangle \nonumber \\
    =& \langle [\mu(t+\tau) - \mu(t)]^2 \rangle +
        \langle [\epsilon(t+\tau) - \epsilon(t)]^2 \rangle  \nonumber \\
        &+2\langle [\mu(t+\tau) - \mu(t)][\epsilon(t+\tau) - \epsilon(t)] \rangle \nonumber \\
    =& S_2^\mu(\tau) + S_2^\epsilon(\tau) + 4R_{\mu\epsilon}(0) - 4R_{\mu\epsilon}(\tau),
\end{align}
where 
\begin{align}
    S_2^\mu(\tau)=\langle [\mu(t+\tau) - \mu(t)]^2 \rangle 
\end{align}
and 
\begin{align}
    S_2^\epsilon(\tau)=\langle [\epsilon(t+\tau) - \epsilon(t)]^2 \rangle
\end{align}
are the second-order structure function of $\mu(t)$ and $\epsilon(t)$, 
the cross-correlation function between $\mu(t)$ and $\epsilon(t)$ is
\begin{align}
    R_{\mu\epsilon}(\tau) &= \langle \mu(t+\tau) \epsilon(t) \rangle =\langle \mu(t) \epsilon(t+\tau) \rangle.
\end{align}

Since $\mu(t)$ and $\epsilon(t)$ are independent of each other,  $R_{\mu\epsilon}(\tau) = 0$. The second-order energy structure function can ultimately be simplified to
\begin{align}
    S_2(\tau) &= S_2^\mu(\tau) + S_2^\epsilon(\tau) \nonumber \\
    &= 2\left(\sigma_\mu^2 + \sigma_\epsilon^2\right) - 2\sigma_\mu^2 R_\mu(\tau) \quad \left( \sigma_\epsilon^2 \ll \sigma_\mu^2 \right) \nonumber \\
    &\approx 2\sigma_\mu^2 - 2\sigma_\mu^2 R_\mu(\tau) \nonumber \\
    &\approx 2\sigma_{_E}^2 - 2\sigma_{_E}^2 R_\mu(\tau).
    \label{eq10}
\end{align}
where $\sigma_\mu$, $\sigma_\epsilon$ and $\sigma_E$ are the standard deviation of $\mu(t)$, $\epsilon(t)$ and $E(t)$, $R_\mu(\tau)$ is the auto-correlation function of $\mu(t)$. Assuming $\epsilon(t)$ represents an uncorrelated random fluctuation (e.g., white noise), the auto-correlation of $\mu(t)$ characterizes the auto-correlation of the event energy. We find that the auto-correlation of both stationary and non-stationary processes can be associated with the second-order structure function in the same form. The second-order structure function has an excellent general applicability when analyzing unknown time series, which is something that auto-correlation functions do not have.

In order to compare energy structure functions between FRBs and earthquakes, we standardize the first and second order energy structure functions respectively using the mean and variance of the energy.
\begin{align}
    N_2(\tau)=\frac{S_2(\tau)}{\sigma_{_E}^2}.
\end{align}
\begin{align}
    N_1(\tau)=\frac{S_1(\tau)}{\mu_{_E}}.
\end{align}

Uncertainties in the structure function are estimated using a bootstrap resampling method. We infer the distribution of energy structure function values within each bin through multiple incomplete sampling of $(\Delta t, \Delta E)$. This enables us to estimate their uncertainties.

\section{Results}
\label{sec:3}

\begin{figure*}
    \centering
    \includegraphics[width=\textwidth]{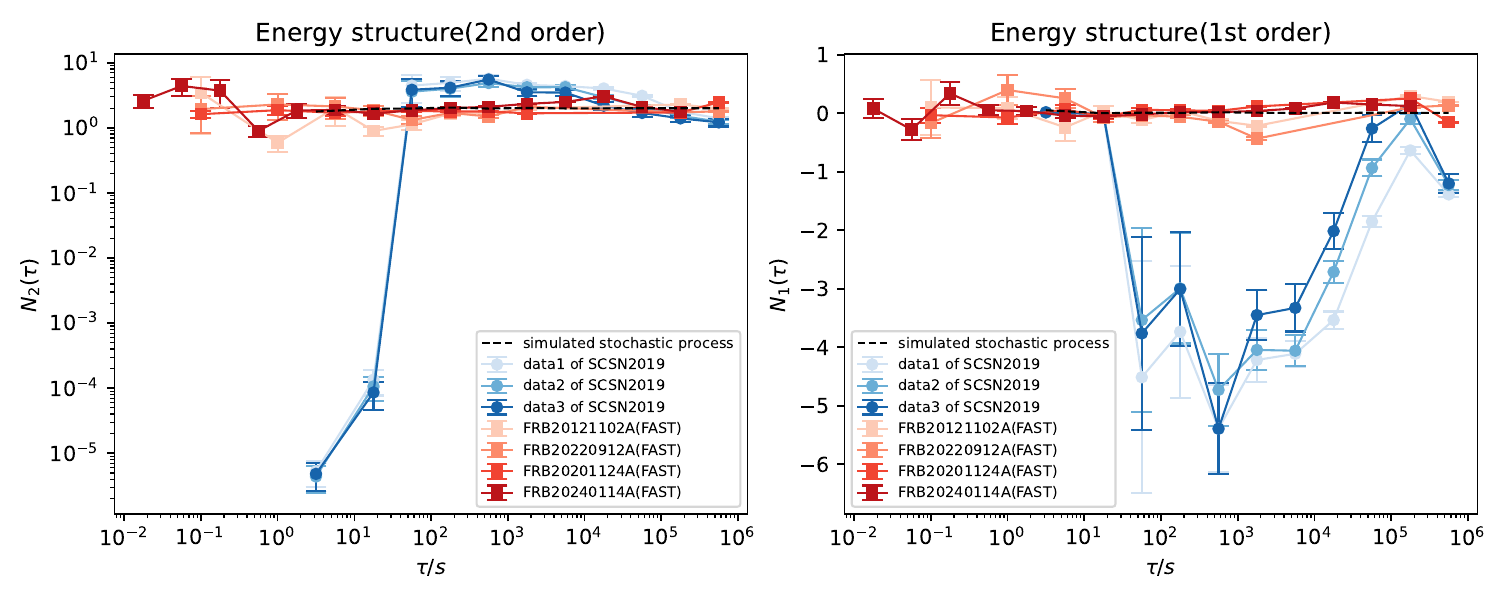}
    \caption{Second-order (left) and first-order (right) energy structure functions as a function of time interval for FRBs, stochastic events, and SCSN earthquakes. Blue lines represent earthquakes, red lines represent FRBs, and the black dashed line corresponds to stochastic events.}
    \label{fig:SCSN}
\end{figure*}

The second-order energy structure functions of all datasets are shown in left panels of Figure \ref{fig:SCSN} and Figure \ref{fig:USGS}. The second-order energy structure functions of seismic datasets increase and then asymptotically approach a constant value. In contrast, the energy structure function of FRBs and Stochastic data remains constant over the entire time scale. 

The second-order energy structure function of earthquakes shows that the energy released by neighboring earthquakes is statistically closer, which is similar to the conclusion of the previous research \citep{Lippiello2008}. In contrast, this relationship does not exist for observed FRBs on time scales ranging from milliseconds to days. This relationship does not exist for stochastically generated seismic data. 

\begin{figure}
    \centering

    \begin{minipage}{\columnwidth}
        \centering
        \captionof{table}{SRM of different datasets}
        \begin{tabular}{cc}  
            \toprule  
            Data Set & SRM\\ 
            \midrule  
            FRB20121102A & 0.028 \\
            FRB20201124A & 0.012 \\  
            FRB20220912A & 0.050 \\ 
            FRB20240114A & 0.019 \\
            data1 of SCSN2019 & 10.385 \\  
            data2 of SCSN2019 & 7.742 \\  
            data3 of SCSN2019 & 6.972 \\  
            data1 of USGC2011 & 6.660 \\  
            data2 of USGC2011 & 6.707 \\  
            data3 of USGC2011 & 6.836 \\  
            Stochastic event & 0.003 \\  
            \bottomrule  
        \end{tabular}  
        \label{tab:SRM}
    \end{minipage}

     \vspace{1em}

     \begin{minipage}{\columnwidth}
         \centering        \includegraphics[width=\textwidth]{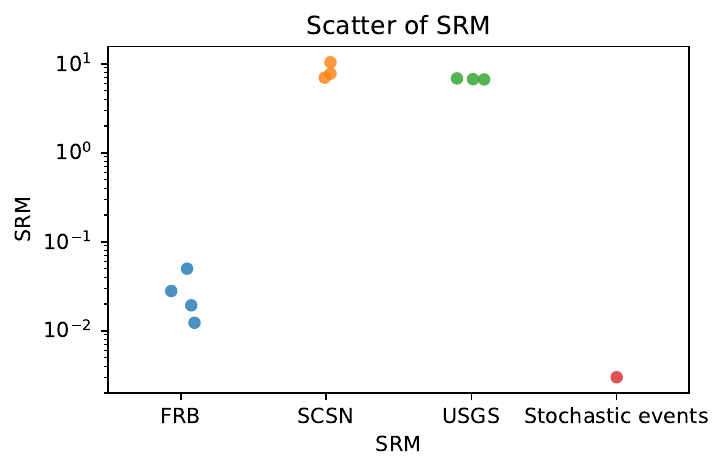}
        \captionof{figure}{Strip plot of the SRM values for different categories. Blue points are FRBs, yellow points are earthquakes in SCSN, green points are earthquakes in USGS and the red point is stochastic event.} 
        \label{fig:srm}
     \end{minipage}
    
\end{figure}

As can be seen from right panels of Figure \ref{fig:SCSN} and Figure \ref{fig:USGS}, similarly, the first-order structure function of seismic datasets exhibits a decreasing and then increasing trend towards zero. In contrast, FRBs and Stochastic events still exhibit a different statistical behavior compared to earthquakes. Their function values stabilize around zero. By introducing the second raw moment (SRM) of the first-order structure function, we are able to clearly measure the extent to which different datasets deviate from the Stochastic process. SRM is defined by 
\begin{align}
    \text{SRM} = \mu_{N_1}^2 + \sigma_{N_1}^2 ,
\end{align}
where $\mu_{_{N_1}}$ is the mean and $\sigma_{_{N_1}}$ is the standard deviation of function values of $N_1$ in all bins. The greater the deviation of the first-order structure function from $N_1(\tau)=0$, the larger the SRM. The SRM of different datasets is shown in Table \ref{tab:SRM}. The similarity between FRBs and Stochastic events and the difference between FRBs and earthquakes can be seen clearly in Figure \ref{fig:srm}. 

The first-order energy structure function shows that there is significant energy attenuation between short-interval earthquakes. This attenuation trend is less significant between earthquakes with longer intervals as the function value gradually rises to zero, which indicates that the energy attenuation is disappearing. Significantly different from earthquakes, there is no obvious energy attenuation or increase in the observed FRBs, which is seen in the Stochastic events.

\section{Analysis}
\label{sec:anl}

\subsection{Data selection}
\label{sec:sel}

The statistical behavior of an earthquake sequence may depend on the duration of the selected time window, as the aftershock rate typically decays with time and may influence the estimation of the energy structure function. To examine the extent to which the selection of the time span influences the conclusions, we conducted a robustness test using datasets of different durations.

\begin{figure*}
    \centering
    \includegraphics[width=\textwidth]{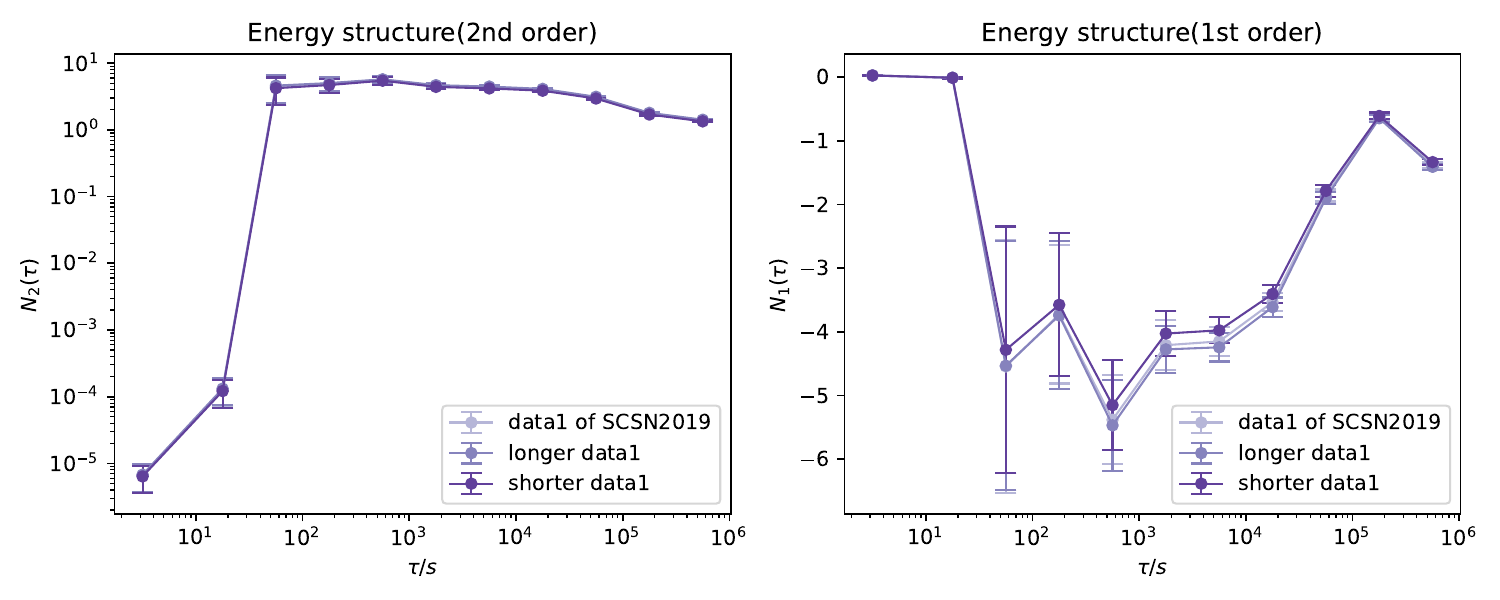}
    \caption{Second- (left) and first-order (right) energy structure functions for earthquakes in SCSN versus time interval. Lines of different shades indicate the different lengths of earthquake series.}
  \label{fig:chose}
\end{figure*}

Specifically, we constructed two additional datasets by varying only the time span while keeping all other parameters identical to those in data1, whose duration is $6\times10^6$s (about 70 days) after the main shock. We separately selected time spans of $4\times10^6$s and $8\times10^6$s and stored them in short and long data sets respectively. As illustrated in Figure \ref{fig:chose}, both the first- and second-order energy structure functions derived from these datasets exhibit nearly identical behaviors to that of data1. This consistency demonstrates that the choice of time scale has little effect on the statistical behaviour of the energy structure function, confirming the robustness of our results.

\subsection{Causality}

\begin{figure}
    \centering
    \includegraphics[width=0.48\textwidth]{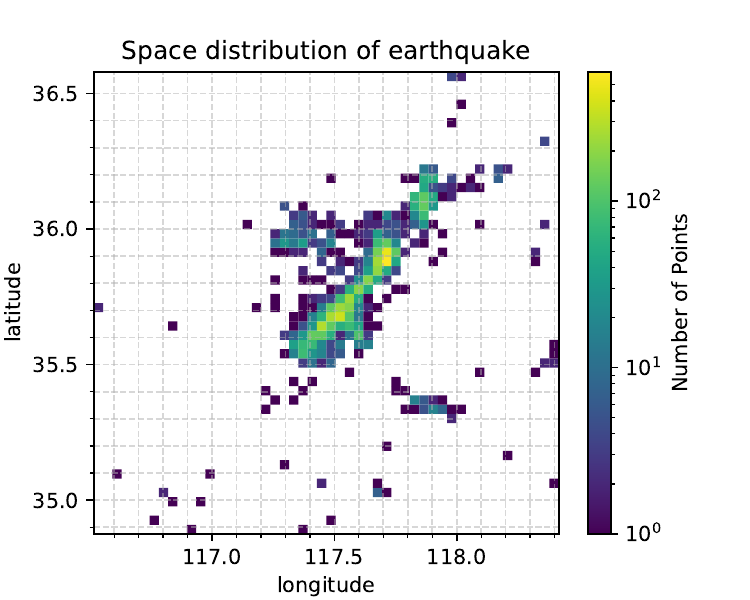}
    \caption{Spatial distribution of earthquakes. The horizontal and vertical axes are longitude and latitude, respectively. The brightness of each area indicates the number of earthquakes occurring within that region. A one-degree difference on the meridian corresponds to 110 km. The distance along the latitude line needs to take into account a factor of cos$\theta_{\text{lat}}$, which is approximately to be $\frac{\sqrt{3}}{2}$. Data sourced from SCSN2019.}
    \label{fig:space}
\end{figure}

Before discussing the results of the energy structure function, we first analyze the causality of the earthquake dataset. The spatial distribution of earthquakes is shown in Figure \ref{fig:space} and Figure \ref{fig:spaceusgs}. Due to spatial size limitations in the selected area, the maximum causal time between earthquakes can be regarded as the longest possible interval between two events that could be connected by the same seismic wave, i.e., the maximum time required for a seismic wave to traverse the entire region. As can be seen in the figure, this seismic zone is distributed along the diagonal of the region. The latitude and longitude of the region span two degrees. Seismic waves travel at speeds of about a few kilometers per second. Thus the maximum causal time can be estimated as
\begin{align}
    t_{\text{cau}} \sim \frac{2\times\sqrt{1^2+ \left( \frac{\sqrt{3}}{2} \right)^2} \times 110 \, \text{km}}{4 \, \text{km/s}}\sim 73 \, \text{s}.
\end{align}

When analyzing the reasons for these differences in energy structure function, we note that the inflection point of the energy structure function of an earthquake is located near 100 seconds, a value within the same order of magnitude to the maximum causal time of this earthquake. It seems to imply that the statistical behavioral characteristics of earthquakes are attributable to correlations between causally linked events. The energy of the seismic wave decays during its propagation through the seismic zone. The energy released when triggering the posterior earthquake is lower than that when triggering prior ones. In the sequence of earthquakes triggered by the same seismic wave, the energy difference is larger for triggered earthquakes with longer intervals since the seismic wave loses more energy. This explains the increase in the second-order energy structure function. For earthquakes with time intervals larger than the maximum causal time, the causal link between them has been lost. Thus the behavior of the energy structure function of earthquakes on the right-hand side reflects the characteristics of independent events. Both FRBs and Stochastic events share similar characteristics over the entire time scale. It implies that FRBs are generated by an independent and Stochastic physical process that is different from earthquakes.

Our conclusions can also be supported by analyzing the auto-correlation. From Eq \ref{eq4} and Eq \ref{eq10}, it can be seen that the second-order structure function reflects the auto-correlation of events for both stationary and non-stationary processes. The auto-correlation function can be expressed in a unified form as
\begin{align}
    R(\tau) &=1-\frac{S_2(\tau)}{2\sigma^2} \nonumber \\
            &=1-\frac{N_2(\tau)}{2}
\end{align}

The closer the auto-correlation function is to zero, the weaker the auto-correlation of the event. The mean values of the auto-correlation function for different data are shown in Table \ref{tab:ACF}. The auto-correlation function for earthquakes decreases rapidly with $\tau$ from 1 to -1, then slowly converges to zero and ends up around 0.3. This also illustrates the weaker auto-correlation between events of FRBs compared to earthquakes.

\begin{table}
  \centering
  \caption{Auto-correlation function (ACF) for different data sets.}
  \label{tab:ACF}
  \begin{tabular}{@{}c c@{}}
    \toprule
    \textbf{Data set name} & \textbf{ACF} \\
    \midrule
    FRB20121102A  &  0.11664  \\
    FRB20201124A  &  0.08962  \\
    FRB20220912A  &  0.08343  \\
    FRB20240114A  & -0.15544  \\
    Stochastic event & 0.00077 \\
    \bottomrule
  \end{tabular}
\end{table}

\subsection{Statistical completeness}
\label{sc}

So far, we have identified the differences between FRBs and earthquakes and the similarities with stochastic process. In order to strengthen the validity of our conclusions, we need to discuss statistical completeness.

The first issue we address is energy completeness. Due to the sensitivity limitations of radio telescopes, observations of FRB sequences are inherently incomplete—only bursts above a certain energy threshold can be detected. This observational bias must be considered when comparing FRBs to earthquakes. Additionally, if FRBs are indeed triggered by starquakes, as some models suggest, it is plausible that weaker starquakes may fail to generate detectable radio bursts at all \citep{Lu2020}, further contributing to incompleteness at the low-energy end.

To account for these factors, we imposed varying minimum energy thresholds on the earthquake datasets (data1–3) to simulate detection incompleteness similar to that affecting FRB observations. Likewise, we applied different energy cutoffs to the FRB data from FRB 20240114A, the most active repeater among the four sources in our sample. By selecting energy thresholds of $10^{36}$, $10^{37}$, and $10^{38}$ erg, we effectively mimic the observational limitations of telescopes with lower sensitivity. As expected, increasing the energy threshold leads to a reduced number of retained bursts, approximating the data quality that might result from less sensitive instruments.

\begin{figure*}
    \centering
    \includegraphics[width=\textwidth]{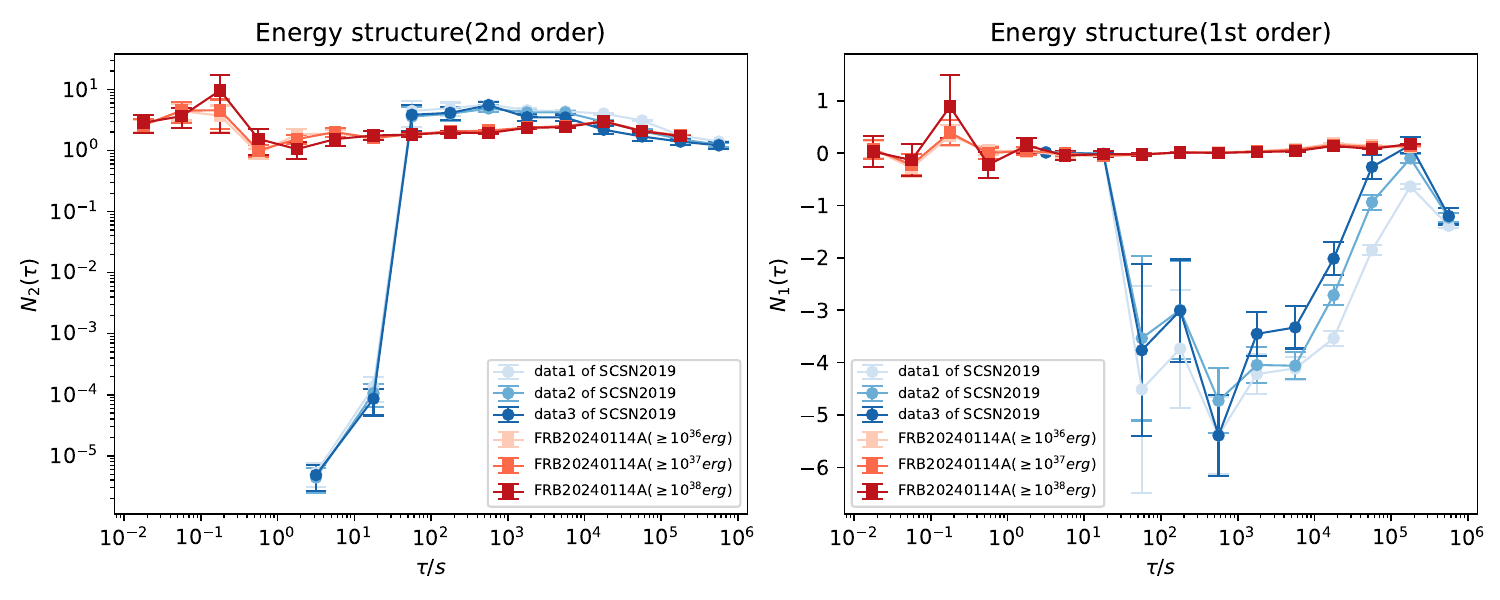}
    \caption{Second- (left) and first-order (right) energy structure functions for FRBs and SCSN earthquakes versus time interval. Blue lines show earthquakes, red lines are FRBs with different energy threshold.}
    \label{fig:SCSN-240114}
\end{figure*}

Despite the reduction in sample size, the energy structure function remains robust: its characteristic behavior is preserved across different threshold settings for both FRBs and earthquakes. This invariance, illustrated in Figure~\ref{fig:SCSN-240114} and Figure~\ref{fig:USGS-240114}, suggests that energy incompleteness does not significantly impact the analysis of the energy structure function, thereby reinforcing its utility as a comparative statistical tool.

The next aspect we discuss is temporal completeness. There is a possibility that the FRBs detected so far are sequences missing the so-called main burst. 

\begin{figure*}
    \centering
    \includegraphics[width=\textwidth]{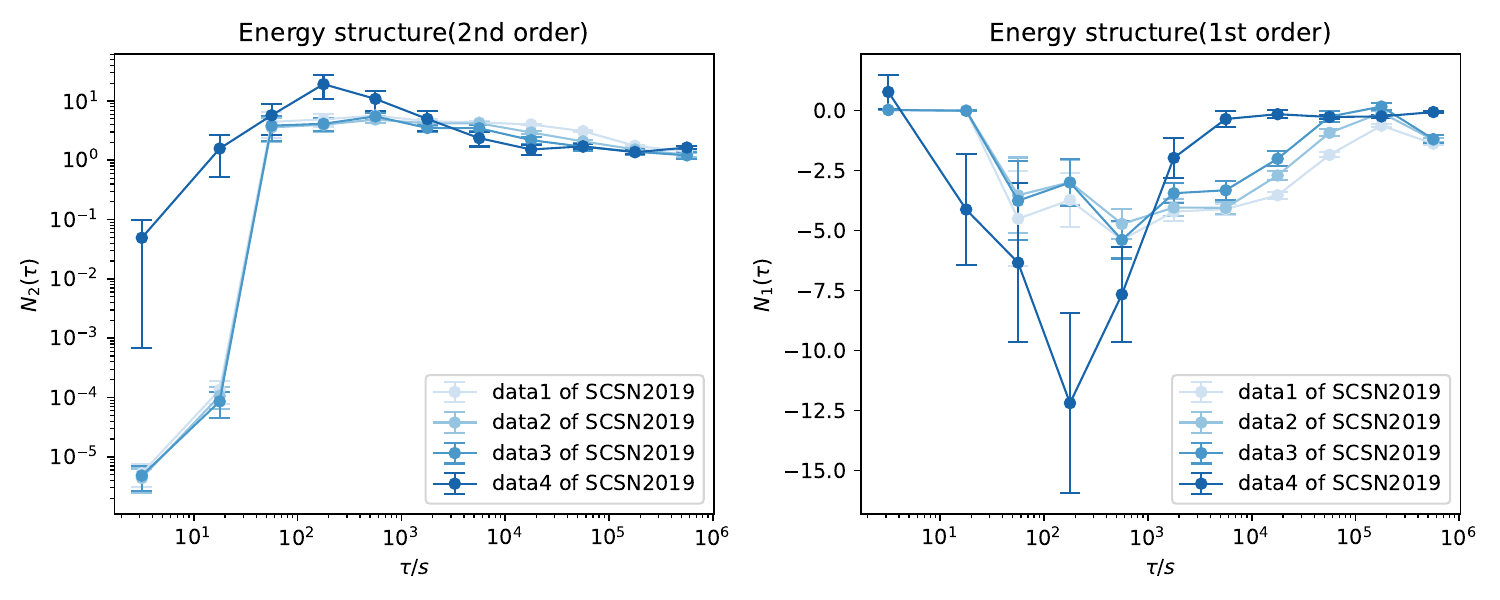}
    \caption{Second- (left) and first-order (right) energy structure functions for earthquakes in SCSN versus time interval. Lines of different shades indicate different datasets.}
    \label{fig:SCSNcomplete}
\end{figure*}

\begin{figure}
    \centering
    \includegraphics[width=0.48\textwidth]{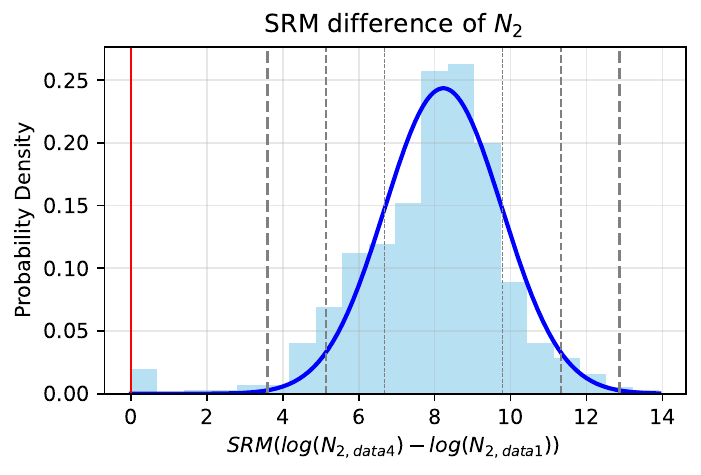}
    \caption{Difference between the second-order energy structure functions of data1 and data4 in SCSN2019. The grey dashed lines represent deviations of 1$\sigma$, 2$\sigma$, and 3$\sigma$ respectively.}
    \label{fig:SRM_difference}
\end{figure}

The flat energy structure function observed in FRBs may result from the non-detection of their main burst. As shown in the left panel of Figure \ref{fig:SCSNcomplete}, compared to earthquake sequences that include a mainshock, the second-order structure function becomes noticeably flatter when the mainshock is absent (data4). To quantify this effect, we repeatedly sampled both the incomplete (data4) and complete (data1) sequences within their respective uncertainties, computed the differences between the two energy structure functions, and evaluated the distribution of the SRM. If the two sequences were statistically consistent, the SRM differences would fluctuate around zero. However, as illustrated in Figure \ref{fig:SRM_difference}, the deviation exceeds the 3$\sigma$ level, confirming that the absence of a mainshock indeed leads to a flattening of the second-order energy structure function.

In contrast, the first-order structure function (right panel of Figure \ref{fig:SCSNcomplete} does not flatten when the mainshock is missing. We performed $10^6$ samplings of the incomplete sequence (data4) within its uncertainty range and found that in $100\%$ of cases, the sampled SRM values were larger than the average SRM of the four FRBs. This demonstrates that even without the mainshock, the first-order energy structure function of earthquakes does not become as flat as that of FRBs. Therefore, the flat energy structure function observed in FRBs is likely an intrinsic property of their emission process, rather than a consequence of missing the main burst.

\section{Conclusions and discussion}
\label{sec:5}

The trigger mechanism of FRBs has been studied for a long time and remains unknown.  Previous studies on time correlations of repeating FRBs have been mainly conducted using the distribution of waiting times, which are time intervals between two successive events. The distribution of waiting times is suggested to be bimodal. The long-waiting time side can be described as a Poisson process \citep{Cruces2021,Hewitt2022,Jahns2023}, while the short-waiting-time component may involve additional physical correlations. To probe this, \citet{Totani2023} introduced a correlation function to explore the physical mechanism related to the short-waiting time side. Different from previous studies on waiting times, they considered the intervals between any two events, not only successive pairs. They proposed that the correlations may exist across all bursts, which is consistent with our consideration. They found that the time correlation function of FRBs deviates from the Poisson process as does that of earthquakes. Some mechanisms are believed by them to be shared by FRBs and earthquakes which form this feature. However, although both FRBs and earthquakes exhibit deviations from the Poisson process on the short-interval side, the specific behaviors of deviation are different \citep{Zhang2024}. The difference is clear in the waiting time distribution and the $\Delta t-\Delta$logE plot of pairs in \citet{Totani2023}. Although we believe that this does not prove that the mechanisms of FRBs and earthquakes are the same, the deviation on the short-interval side from the Poisson process is a question worth considering. By setting different energy thresholds to filter the FRB data, it was found that the left peak (short-waiting time) of the waiting time distribution is insensitive to the energy threshold, while the right peak (long-waiting time) shows a strong dependence \citep{Zhang2024}. As the energy threshold increases, the events become sparser in time. The rightward shift of the peak of the waiting time distribution is reasonable. However, there is no obvious change in the left peak. This suggests that there might be some intrinsic connection between the bursts with short intervals. The geometric configuration of the radiation cone may explain this \citep{Beniamini2025ApJ}. For instance, Cherenkov radiation is emitted in a characteristic hollow cone \citep{Liu2023}. Under this scenario, we can merge the bursts of short waiting times. The peak on the right is the waiting time distribution of the triggering events. This indicates that the triggering events are stochastic.

In addition to the analysis of the time dimension, energy can also be served as a perspective for studying the triggering mechanism of FRBs. By comparing the conditional probability that the energy of posterior events is lower than that of prior events before shuffling and after shuffling, it is proved that there is no trend of energy attenuation in FRBs, which is consistent with the conclusion we obtained through the first-order energy structure function. The research on the Pincus Index further provides evidence that the origin of FRBs is highly Stochastic \citep{Zhang2024}. The high randomness and low chaos of FRBs make them more similar to Brownian motion than to earthquakes. This method also shows that FRBs are different from magnetar bursts, which suggests that FRBs may not originate from typical magnetar bursts \citep{Yamasaki2024MNRAS}.

Based on the above analysis, the evidence appears to favor the view that FRBs may not emit bursts in the same way as earthquakes. The source of this difference lies in the correlation of the trigger mechanism of earthquakes and the stochasticity of the trigger mechanism of FRBs. Although the starquake model can well explain some of the statistical results, it is still difficult to determine whether FRBs originate from starquakes or not. This is because the conformity of the statistical results is a necessary but insufficient condition. The analysis of the core contradiction is missing. 

By comparing the energy structure function of FRBs with those of earthquakes and stochastic process, we found that their statistical behavior exhibit the following characteristics. First, we revealed the similarities between FRBs and earthquakes on the long-interval side and the differences on the short-interval side. We suggested that FRBs are more likely a series of independent events, because the causal chains in earthquakes produce their distinct behavior across timescales. Second, the similarities in first-order energy structure functions of FRBs and a stochastic process indicate that there is neither energy decay nor increase in the FRB sequence. The similarities in second-order energy structure functions of FRBs and a stochastic process imply that FRBs may originate from diverse emission mechanisms or sites. This makes the difference in energy between two bursts independent of the interval. Additionally, we discuss the statistical completeness to enhance the credibility of our conclusions. We demonstrate that energy incompleteness will not smooth out the characteristic features of earthquakes in the energy structure function. We also ruled out temporal incompleteness, since the distribution of SRM of the first-order energy structure function for FRBs and the missing mainshock earthquake sequence exhibits a marked difference.

However, there remain some questions regarding this topic that require further discussions. First, since the magnetar environment differs from that of Earth, the dynamical timescales for magnetar quakes and earthquakes may be different. The shorter characteristic timescale on magnetars may cause FRBs to exhibit correlations at smaller timescales. Given the temporal precision of current observations, we can only confirm that no correlations exist for FRBs at timescales exceeding 10 milliseconds. Second, the randomness we find may also result from a selection effect. It is possible that only a part of the starquakes satisfying specific conditions produced the FRBs we observed. This interpretation could also account for the randomness of FRBs. However, it implies that the number of bursts we observe is smaller than the number of actual bursts. A further discussion is needed on how this mechanism works to produce FRBs with a high burst rate.

By analyzing the energy structure function of FRBs, we found that there is no trend of increasing or decreasing in the energy of FRBs and that the differences between events do not vary with time interval scale. This provides supporting for the stochastic origin model of FRBs.

\section*{Acknowledgements}

We thank the referee for valuable suggestions that have allowed us to improve the quality of the manuscript. We acknowledge valuable discussion with Sen-Lin Pang, Yu-Chen Huang, Jia-Pei Feng and Le-Yan Shen. This work made use of data from FAST, a Chinese national mega-science facility built and operated by the National Astronomical Observatories, Chinese Academy of Sciences, and SCSN catalog of earthquakes developed by the California Institute of Technology and the United States Geological Survey Pasadena. This work was supported by  the National Natural Science Foundation of China (grant No. 12393812), the National SKA Program of China (grant No. 2020SKA0120302), and the Strategic Priority Research Program of the Chinese Academy of Sciences (grant NO. XDB0550300). Yong-Kun Zhang is supported by the Postdoctoral Fellowship Program and China Postdoctoral Science Foundation under Grant Number BX20250158.

\section*{Data availability}
The data of FRB 20121102A are available in \citet{Li2021}.
The data of FRB 20201124A are available in \citet{FRB201124}. The data of FRB 20220912A are available in \citet{FRB220912}. 
The data of FRB 20240114A are available in \citet{Zhang2025arXiv}. The data of earthquakes in SCSN are available in \url{https://www.fdsn.org/networks/detail/CI/}. The data of earthquakes in USGS are available in \url{https://www.usgs.gov/natural-hazards/earthquake-hazards/lists-maps-and-statistics}.

\bibliographystyle{mnras}
\bibliography{ms} 

\appendix
\setcounter{figure}{0}
\renewcommand{\thefigure}{A\arabic{figure}}
\section*{Appendix: figures of detailed results for the data sets not presented in the main part}

Figures relating to the USGS dataset are presented here.

\begin{figure*}
    \centering
    \includegraphics[width=\textwidth]{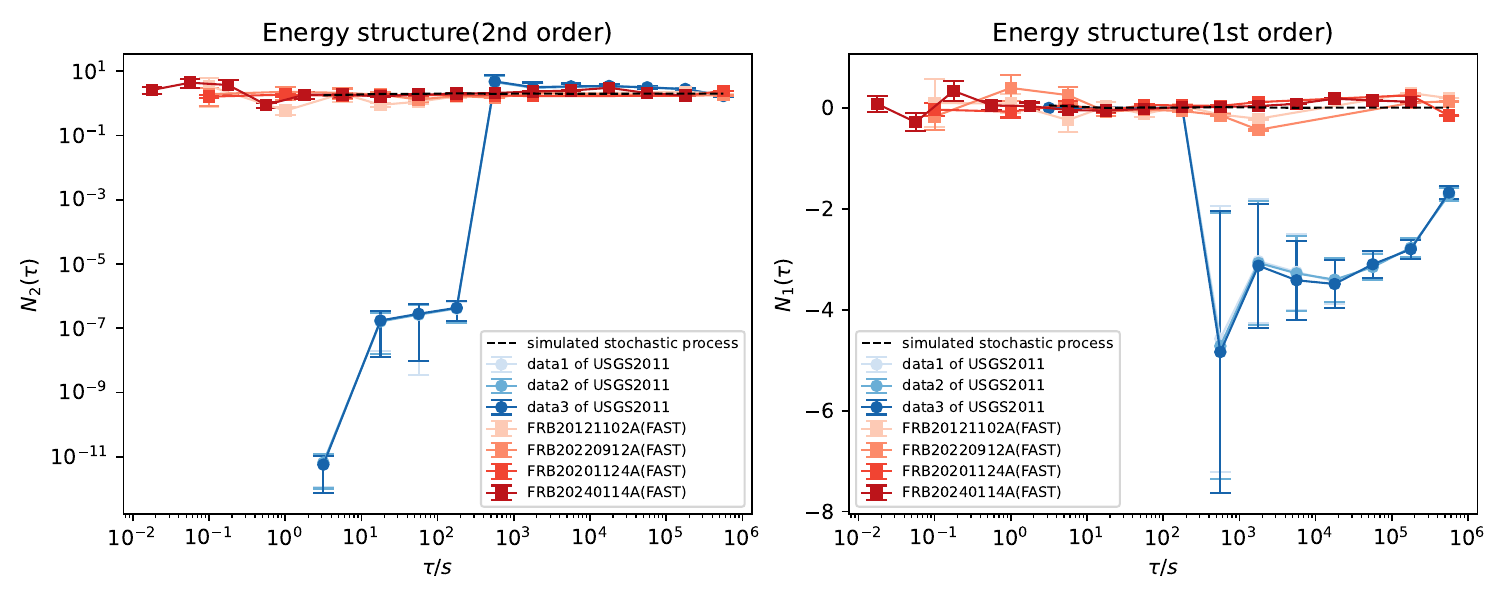}
    \caption{Similar to Figure \ref{fig:SCSN} but using data of earthquakes from USGS instead of SCSN.}
    \label{fig:USGS}
\end{figure*}

\begin{figure*}
    \centering    \includegraphics[width=\columnwidth]{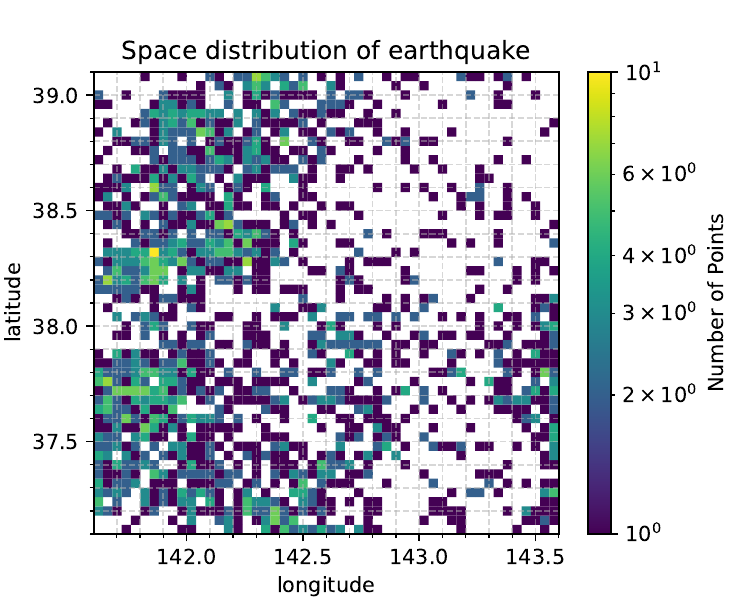}
    \caption{Spatial distribution of earthquakes. Data sourced from USGS2011.}
    \label{fig:spaceusgs}
\end{figure*}

\begin{figure*}
    \centering
    \includegraphics[width=\linewidth]{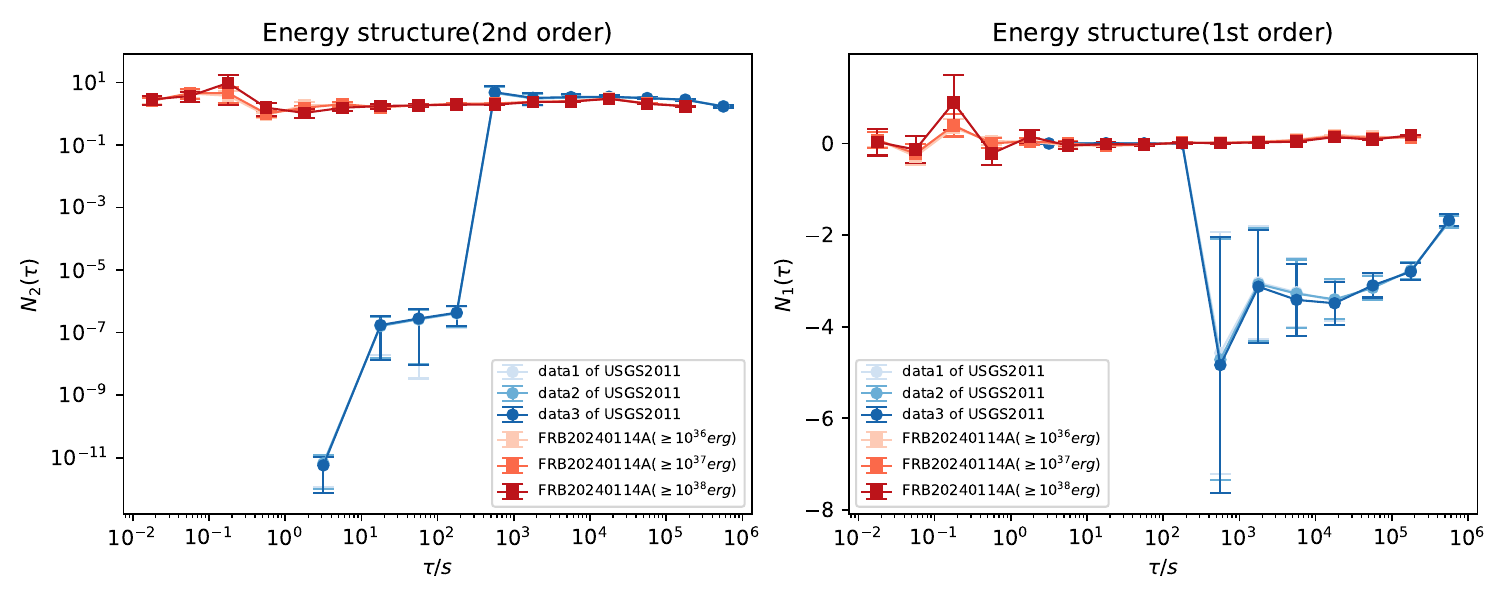}
    \caption{Similar to Figure \ref{fig:SCSN-240114} but using data of earthquakes from USGS instead of SCSN.}
    \label{fig:USGS-240114}
\end{figure*}

\bsp	% typesetting comment
\label{lastpage}
\end{document}